\documentclass[aps,prd,a4paper,superscriptaddress,nofootinbib,10pt, two column]{revtex4-1}
\usepackage{amsmath,amssymb,graphicx,multirow,dcolumn,bm,latexsym,soul,nicefrac}
\usepackage{epstopdf}
\usepackage{mathrsfs}
\usepackage{acronym}
\usepackage[colorlinks,linkcolor=blue,citecolor=blue,urlcolor=blue ]{hyperref}
\usepackage{ulem}
\usepackage{subcaption}
\usepackage{float}
\usepackage{xcolor}
\usepackage[labelformat=simple,labelsep=period,skip=3pt]{caption}
\newcounter{RomanNumber}

\allowdisplaybreaks
\newcommand{\be}{\begin{equation}}
\newcommand{\ee}{\end{equation}}
\newcommand{\bea}{\begin{eqnarray}}
\newcommand{\eea}{\end{eqnarray}}
\newcommand{\nn}{\nonumber}

\newcommand{\msun}{{\rm M}_\odot}

\newcommand{\TRC}{TianQin Research Center for Gravitational Physics and School of Physics and Astronomy, Sun Yat-sen University (Zhuhai Campus), Zhuhai 519082, People's Republic of China}
\newcommand{\CGE}{MOE Key Laboratory of Fundamental Physical Quantities Measurement and Hubei Key Laboratory of Gravitation and Quantum Physics, PGMF and School of Physics, Huazhong University of Science and Technology, Wuhan 430074,  People's Republic of China.}

\begin{document}

\title{Science with the TianQin observatory: Preliminary result on extreme-mass-ratio inspirals}

\author{Hui-Min Fan}
\affiliation{\CGE}

\author{Yi-Ming Hu}
\email{huyiming@mail.sysu.edu.cn}
\affiliation{\TRC}

\author{Enrico Barausse}
\affiliation{SISSA, Via Bonomea 265, 34136 Trieste, Italy, and INFN Sezione di Trieste, via Valerio 2, 34127 Trieste, Italy}
\affiliation{IFPU --Institute for Fundamental Physics of the Universe, Via Beirut 2, 34014 Trieste, Italy}
\affiliation{Institut d'Astrophysique de Paris, CNRS and Sorbonne
 Universit\'es, UMR 7095, 98 bis bd Arago, 75014 Paris, France}

\author{Alberto Sesana}
\affiliation{Dipartimento di Fisica ``G. Occhialini", Universit\'{a} degli Studi Milano Bicocca, Piazza della Scienza 3, I-20126 Milano, Italy}

\author{Jian-dong Zhang}
\email{zhangjd9@mail.sysu.edu.cn}
\affiliation{\TRC}

\author{Xuefeng Zhang}
\affiliation{\TRC}

\author{Tie-Guang Zi}
\affiliation{\TRC}

\author{Jianwei Mei}
\email{meijw@sysu.edu.cn}
\affiliation{\TRC}

\date{\today}

\begin{abstract}\label{sec:abs}

Systems consisting of a massive black hole and a stellar-origin \ac{CO}, known as extreme-mass-ratio inspirals (EMRIs),
are of great significance for space-based gravitational-wave detectors, as they will allow for testing gravitational theories
in the strong field regime, and for checking the validity of the black hole no-hair theorem. In this work, we present a calculation of 
the EMRI rate and parameter estimation capabilities of the TianQin observatory, for various astrophysical models for these sources. We find 
that TianQin can  observe EMRIs  involving COs with  mass of 10$M_\odot$ 
up to redshift  $\sim2$.
We also find that detections could reach tens or hundreds per year in the most optimistic astrophysical scenarios. Intrinsic parameters are expected to be recovered to within
fractional errors of  $\sim 10^{-6}$, while  typical errors on the luminosity distance and sky localization are
10\% and  10 deg$^2$, respectively. 
  TianQin observation of EMRIs can also constrain possible deviations from the Kerr quadrupole moment to within fractional errors $\lesssim10^{-4}$.
  We also find that a network of multiple detectors would allow for improvements  in both detection
  rates (by a factor $\sim 1.5$ -$3$) and in parameter estimation precision (20-fold improvement for the sky localization and fivefold improvement for the other parameters).

\end{abstract}

\maketitle
\acrodef{PSD}{power spectral density}
\acrodef{SNR}{signal-to-noise ratio}
\acrodef{EMRI}{extreme mass ratio inspiral}
\acrodef{FIM}{Fisher information matrix}
\acrodef{GW}{Gravitational-wave}
\acrodef{MBH}{massive black hole}
\acrodef{CO}{compact object}
\acrodef{pc}{parsec}
\acrodef{LSO}{last stable orbit}
\acrodef{AK}{analytic kludge}
\acrodef{NK}{numerical kludge}
\acrodef{AAK}{augmented analytic kludge}

\section{Introduction}\label{sec:intro}


\ac{GW} observations provide information on the minute vibrations of the spacetime, 
and promise to revolutionize astronomy and astrophysics by opening a new window on the Universe.
To date, the ground-based GW observatories, LIGO and Virgo, have detected several GW events \cite{LIGOScientific:2018mvr,Abbott:2020uma,LIGOScientific:2020stg,Abbott:2020khf}.
Limited by seismic noise and their relatively short arm lengths, however, ground-based detectors are  sensitive  only to high-frequency GWs (above a few hertz) generated by low-mass sources [e.g., mergers of stellar-origin compact objects(COs)].
In order to detect heavier sources, such as ones involving \ac{MBH}, or even the low-frequency (subhertz) inspiral phase
of stellar-origin compact binaries \cite{Sesana:2016ljz},  a significant increase in the size of GW detectors  is necessary, which
can  be achieved only in space.
LISA, for example, will present arm lengths of about 2.5 million km and will be sensitive to GWs in the frequency band $10^{-5}$ -$0.1$~Hz \cite{Audley:2017drz,Baker:2019nia}.


TianQin is a proposed space-based, geocentric \ac{GW} observatory
with arm lengths of about $1.7\times10^5$ km, aiming to detect \ac{GW} signals in the frequency band $10^{-4}$ -$1$~Hz \cite{Luo:2015ght,Hu:2019}. In the past few years, a systematic effort has been undertaken to study the science prospects of TianQin \cite{Hu:2017}. On the astrophysics side, this included the
study of the detection prospects for Galactic ultracompact binaries \cite{Huang:2020rjf},  coalescing \acp{MBH} \cite{Wang:2019ryf,Hu:2018yqb}, the low-frequency inspiral of stellar-mass black holes \cite{Liu:2020eko}, and stochastic \ac{GW} backgrounds \cite{Liang:2019}. On the fundamental physics side, 
TianQin's ability to test the black hole no-hair theorem with the ringdown of \acp{MBH} resulting from a merger has been analyzed,
both in a theory-agnostic framework \cite{Shi:2019hqa} and within specific gravitational theories extending general relativity \cite{Bao:2019kgt},
and more work is in preparation in this direction.

In this paper, we focus on extreme-mass-ratio insprials (EMRIs), i.e., binaries consisting of a stellar-origin \ac{CO} (a stellar mass
black hole or a neutron star) orbiting around a \ac{MBH} in a long inspiral \cite{AmaroSeoane:2007aw,Berry:2019wgg}. These sources are expected to
be relatively numerous in the millihertz GW sky probed by LISA and TianQin. Indeed, strong observational evidence suggests the presence of 
\acp{MBH} at the center of most local galaxies  \cite{kormendy1995inward,magorrian1998demography,Gebhardt:2002js, Ferrarese:2004qr},
typically surrounded by stellar clusters or cusp s\cite{Alexander:2005jz,Schodel:2014wma} of a few parsec scale. 
 Relaxation processes in such a high-density environment occasionally force stars and compact objects onto extremely eccentric, low angular momentum orbits, resulting in a close encounter with the central MBH. While main sequence stars are torn apart and tidally disrupted, potentially resulting in luminous flares and prompting gas accretion onto the central MBH \cite{1975Natur.254..295H,1991ApJ...370...60M,Freitag:2002mj,Gezari:2003wz}, COs typically survive intact until merger \cite{Merritt:2015kba,Bar-Or:2016qop,Amaro-Seoane:2019umn}.
 Depending on  their orbital angular momentum, COs can directly plunge into the MBH or be captured in eccentric bound orbits, whose secular evolution decouples from the cluster's dynamics and is dominated by GW emission \cite{AmaroSeoane:2007aw}. These latter systems are usually referred to as EMRIs.

Detecting \acp{GW} from EMRIs will be very significant for our understanding of the astrophysics of these
sources \cite{AmaroSeoane:2007aw}. For instance, it will allow for gaining information on
the mass distribution of MBHs \cite{Gair:2010yu} and their host stellar
environments \cite{AmaroSeoane:2007aw}.
It may also provide information on the expansion of the Universe \cite{MacLeod:2007jd}, 
as well as allow for mapping the spacetime geometry of the \ac{MBH} in great detail so that a stringent test of general relativity is possible \cite{Gair:2012nm}, through measuring the bumpy nature of black holes \cite{Piovano:2020ooe}, investigating the non-Kerr nature of central objects \cite{Glampedakis:2005cf},  constraining modified gravity theories \cite{Niu:2019ywx},
and testing the no-hair theorem \cite{Ryan:1995wh,Ryan:1997hg,Ryan:1997kh,Barack:2006pq,Chua:2018yng}.
Futhermore, it can also reveal the possible presence of matter surrounding MBHs through their effects on waveforms \cite{Barausse:2006vt,Barausse:2007dy,Gair:2010iv,Yunes:2011ws,Barausse:2014tra,Barausse:2014pra,Derdzinski:2020wlw}.



To estimate the detection rate for EMRIs with a specific detector, one would need the corresponding astrophysical models as input, including the description of the population properties as well as the event rates \cite{Mapelli2012ACV,Berry:2016bit,Bortolas:2019sif}.
In this paper, we use previously published astrophysical models \cite{Babak:2017tow} for the formation and evolution of EMRIs across cosmic time to
assess TianQin's capability to detect these sources and estimate their parameters.
 In more detail,
by adopting, as was done in Ref. \cite{Babak:2017tow} for LISA, analytic kludge waveforms 
and using a simple \ac{FIM} method to analyze the
parameter estimation prospects,
 we find that
EMRIs  can be observed up to redshift  $\sim2$, assuming a 10 $M_\odot$ CO, with  rates vary from  $10$ to 100 yr$^{-1}$.
Intrinsic parameters are projected to be estimated to within
fractional errors of $\sim 10^{-6}$, while typical errors on the luminosity distance and sky localization are
 10\% and 10 deg$^2$, respectively. 
We also estimate the potential scientific gain of operating TianQin within a network of detectors, e.g.,
together with LISA and/or a twin TianQin detector (TQ II). 
We find that a twin TianQin detector
would increase detection rates by a factor of $\sim 1.5$ -$3$.


The paper is organized as following.
In Sec. \ref{sec:back}, we describe our model for the EMRI astrophysical population, the gravitational waveforms, the response of TianQin to \ac{EMRI} signals and the TianQin noise model used in this study.
In Sec. \ref{sec:method}, we describe the method for calculating the \ac{SNR} and the precision of the parameter estimation.
The main results of our study are presented in Sec. \ref{sec:result}.
In Sec. \ref{summary}, we present our conclusions.

\section{The model}\label{sec:back}

\subsection{EMRI rate}\label{subsec:astro}

Extensive evidence exists for the ubiquitous presence of \acp{MBH} at the center of virtually every galaxy at low redshifts  \cite{1971MNRAS.152..461L,Soltan:1982vf,Kormendy:1995er,Gultekin:2009qn}, including our own Milky Way \cite{Schodel:2002py,Reid:1999kt,Reid:2002hw,2018A&A...618L..10G} and, as recently confirmed by the Event Horizon Telescope, M87 \cite{Akiyama:2019cqa}.
Moreover, nuclear stellar clusters with sizes of a few parsecs(pc) and masses up to $10^7$ -${10^8} M_\odot$ are also known to coexist with \acp{MBH} in the local Universe (except at the high-mass end of the \ac{MBH} mass function) \cite{Schodel:2014wma}.
The high densities of these clusters make them the perfect cradles for the formation of \ac{EMRI}s, as two-body relaxation will make the system tend toward energy equipartion and, thus,  mass segregation, with the heavier objects (i.e., stellar-mass black holes) sinking deeper in the \ac{MBH} gravitational potential well.
This process can eventually lead to \acp{CO} plunging or inspiraling into the \ac{MBH}, depending on their angular momenta.


The rate of EMRIs and their properties depend on a variety of (astro)physical processes, which determine the evolution of the population of \acp{MBH} along cosmic history and the accumulation of \acp{CO} in their vicinity.
In this paper, we make use of the EMRI population models developed by Babak {\itshape{et al}}. \cite{Babak:2017tow}.
For the convenience of the readers, we summarize the main features of the model here but refer to Ref. \cite{Babak:2017tow} for more details.
The intrinsic EMRI rate is given by the following function:
\be
{\cal R}(M,z,a)=\frac{d^3N}{dMdzda}\times p_0(M,z)\times\kappa\Gamma{R_0(M)}\,,
\label{formu:distribution}
\ee
where $M,z,$ and $a$ are the mass, redshift, and spin of the \ac{MBH}, respectively. Terms on the right-hand side of this equation are explained below.
\begin{itemize}
\item $d^3N/(dMdzda)$ is the redshift-dependent \ac{MBH} mass function.
    Two scenarios have been adopted that bracket the uncertainties in the low-mass end of the \ac{MBH} mass function at $z=0$.
    One scenario is based on the semianalytic model (SAM) developed by Barausse and collaborators in a series of papers \cite{Barausse:2012fy,Sesana:2014bea,Antonini:2015sza}. The SAM follows the formation and evolution of \ac{MBH} masses and spins along cosmic history, and produces a mass function $dN/d{\rm log}M\propto M^{-0.3}$ in the range $10^5 \msun< M < 10^7 \msun$, consistent with the upper bound of current observations (see Fig. 1 in Ref. \cite{Babak:2017tow}).
    A second scenario employs an empirical mass function where $dN/d{\rm log}M\propto M^{0.3}$ in the same mass range $10^5 \msun< M< 10^7 \msun$ \cite{Gair:2010yu}, consistent with the lower bound of current observations.

\item $p_0(M,z)$ describes the probability that a \ac{MBH} with mass $M$ and redshift $z$ is surrounded by a cusp of stars and \acp{CO}, thus potentially giving rise to an {\it active} EMRI source.
    When two galaxies merge, the cusps of stars and \acp{CO} around the central \acp{MBH} of the parent galaxies are eroded by the action of the inspiraling \ac{MBH} binary.
    The cusp in the merger remnant is, therefore, destroyed, and is rebuilt only after a fraction of the relaxation time \cite{AmaroSeoane:2010bq}.
    The SAM model allows one to follow the galaxy and \ac{MBH} merger rate and to estimate the time needed to rebuild the cusp, from which the probability function for each \ac{MBH} to be a potential EMRI source is constructed.

\item ${R_0(M)}$ is the rate at which a galaxy hosting a MBH with mass $M$ surrounded by a stellar (and CO) cusp actually generates an EMRI. Note that this probability depends on the density profile of the \ac{CO} population, which might depend on the redshift and other parameters unrelated to the \ac{MBH} mass. For simplicity, however, any such possible dependence is dropped, and ${R_0(M)}$ is assumed to be  a function of the \ac{MBH} mass only, following Ref. \cite{AmaroSeoane:2010bq}.

\item Finally, $\kappa$ and $\Gamma$ are two ``{\itshape{ad hoc}}'' correction factors to ${R_0(M)}$ that ensure that the overall EMRI rate is consistent with the observed \ac{MBH} mass function, i.e., that the \acp{MBH} do not ``overgrow`'' their present masses by capturing too many EMRIs and plunges.
\end{itemize}

Besides the choice of two different \ac{MBH} mass functions, Eq. \eqref{formu:distribution} -- and the observed EMRI rate --also depends on a number of additional astrophysical factors, including:

\begin{itemize}
\item The relative occurrence rate of plunges versus EMRIs, which enters in the $\Gamma$ factor.--The plunge cross section of the \ac{MBH} itself ($4GM/c^2$ for a nonrotating \ac{MBH}) is not negligible compared to the EMRI capture cross section, which is generally $<10GM/c^2$.
    Recent simulations have actually found that plunges are typically more frequent than EMRIs  \cite{Merritt:2015kba}, and this has an impact on the intrinsic EMRI rate for a given \ac{MBH}.

\item The choice of parameters of the \ac{MBH}-galaxy scaling relations,  which is important to compute the $p_0(M,z)$ function.--In fact, the time needed to rebuild the cusp depends on the properties of the galactic nucleus, whose mass can be computed from the \ac{MBH} mass via the \ac{MBH}-galaxy scaling relations.

\item The \ac{MBH} spin distribution, which has an impact both on the EMRI capture rate and the EMRI waveforms, and, hence, on their detectability with \acp{GW} (as shown in the following section).

\item The mass of the \ac{CO}, which affects the rate normalization and which enters the EMRI waveforms.--Most models in Ref. \cite{Babak:2017tow} assume \acp{CO} with $10 \msun$, but some consider \acp{CO} with $30 \msun$.
\end{itemize}

The above ingredients have been suitably combined in Ref. \cite{Babak:2017tow} to build a suite of 12 models encompassing three orders of magnitude in the expected cosmic EMRI rate, from 10 to about $2\times10^4$ yr$^{-1}$.
These are also the models that we employ in this investigation.
We label the models as M-$i$ with $i=1,\dots,12$, following the original nomenclature.
Detailed prescriptions for each model can be found in Table I of Ref. \cite{Babak:2017tow}.

\subsection{Waveform}\label{sub:wave}

The calculation of the waveforms for \acp{EMRI} is a challenging task.
Although much progress has been attained with the goal of producing accurate and efficient \ac{EMRI} waveforms including the effect of the self-force~\cite{Drasco:2005kz,Warburton:2011fk,vandeMeent:2016pee,Warburton:2014bya,Huerta:2008gb,vandeMeent:2018rms,Pound:2019lzj}, the problem has not been fully solved yet.
Here, we adopt simple waveforms suitable for predicting the detection and parameter estimation capabilities of TianQin, but one should keep in mind
that full waveforms including the effect of the self-force will be needed to analyze the real data.

In more detail, in this paper we follow Ref. \cite{Babak:2017tow} and utilize a class of
simplified and approximate but computationally inexpensive \ac{EMRI} waveforms, the analytic kludge (AK) model from Ref. \cite{Barack:2003fp}. There are other kludge waveforms available for EMRI events, like the numerical kludge (NK) \cite{Glampedakis:2002cb,Gair:2005ih,Babak:2006uv} or the augmented analytic kludge (AAK) \cite{Chua:2015mua,Chua:2016jnd,Chua:2017ujo}, with NK being significantly slower than AK and AAK being comparable to AK in terms of computing time. 
Both the NK and the AAK waveform are physically more self-consistent compared with the AK waveform, but neither is fully consistent, and for the sake of fair comparison, we stick with the AK waveform for the following computations.

The AK waveform is calculated simply from the quadrupole formula,
while the orbital evolution of the \ac{CO}  includes post-Newtonian (PN) corrections accounting for pericenter precession,  Lense-Thirring precession, and (leading-order) radiation reaction.

The waveform far away from the source is given in the transverse traceless gauge by
\be
h_{ij}=\frac{2}{D}(P_{ik}P_{jl}-\frac{1}{2}P_{ij}P_{kl})\ddot{I}^{kl}\,,
\label{nu_Sch}
\ee
where $D$ is the source distance,
$P_{ij}\equiv\eta_{ij}-\hat{n}_i\hat{n}_j$ is the projection operator
on the space orthogonal to the unit vector of the source position $\hat{n}$,
and $\ddot{I}^{ij}$ is the second time derivative of the quadrupole.
For an EMRI system with CO mass $m$, central MBH mass $M$, and mass ratio $m/M\ll1$, we have $I^{ij}(t)=mr^i(t)r^j(t)$,
where  $\vec{r}$ is the displacement vector of the CO from the MBH.

The orbit evolution of the CO is described by the first-order derivative of the following five quantities:
\begin{itemize}
\item $\Phi$, the mean anomaly of the CO's orbit;
\item $\nu$, the orbital frequency;
\item $e$, the orbital eccentricity;
\item $\alpha$, the azimuthal angle of the CO orbital angular momentum $\vec{L}$ with respect to the MBH's spin angular momentum $\vec{S}$; and
\item $\tilde{\gamma}$, the direction of the pericenter relative to $\vec{L}\times\vec{S}$.
\end{itemize}
The evolution equations for $\Phi$, $\nu$, and $e$ include terms up to  3.5PN order. The evolution of $\alpha$ caused by Lense-Thirring precession and that of $\tilde{\gamma}$ caused by pericenter precession are instead  followed up to  2PN order.
The equations depend on the two masses $m$ and $M$, the dimensionless spin $a=S/M^2$, and the angle $\lambda$ between $\vec{L}$ and $\vec{S}$. For a distant source, the masses should be replaced by ``redshifted'' mass $m_z=m(1+z)$ and $M_z=M(1+z)$.
To test the no-hair theorem, one can also introduce an arbitrary quadrupole moment $Q$ for the central MBH in the evolution equations.
The explicit form of the evolution equations is given in Eqs. (27-31) of Ref. \cite{Barack:2003fp} (without $Q$) and in Eqs. (4-8) of Ref. \cite{Barack:2006pq} (with an arbitrary $Q$).
Obviously,  the orbital evolution depends on the initial conditions at some initial time $t_0$.
To obtain the waveform, we also need five other extrinsic parameters, namely, the source's sky position ($\theta_S$, $\phi_S$),
the luminosity distance $D$, and direction of the MBH's spin $\vec{S}$ relative to the line of sight ($\theta_K$ and $\phi_K$).
In summary, there are 14 parameters,
$(t_0, m, M, a, e_0, \tilde{\gamma}_0, \Phi_0, \theta_S, \phi_S, \lambda, \alpha_0, \theta_K, \phi_K, D_L)$, with the additional
parameter $Q$ introduced when testing the no-hair theorem.

In Ref. \cite{Barack:2003fp}, the waveform was conventionally cut off
at the \ac{LSO} $r_{\rm LSO}$ of the Schwarzschild spacetime. We refer
to this as the AK Schwarzschild (AKS) case. However,
since the exact value of the cutoff frequency can have a significant impact
on  the parameter estimation, we also consider, as in Ref. \cite{Babak:2017tow}, an AK Kerr (AKK) waveform
model where we cut the waveform off at the Kerr LSO. As argued in Ref. \cite{Babak:2017tow}, more realistic 
EMRI waveforms including dissipative self-force effects should yield results
between the (more conservative) AKS results and the (more optimistic) AKK ones.

\subsection{Detector response}\label{sub:dataanal}

TianQin will consist of three satellites orbiting the Earth, forming a regular triangular constellation, with each side measuring about $L=1.7 \times 10^8$ m.
The detector orientation, i.e., the direction normal to the plane of the constellation, will point to a reference source, the white dwarf binary system RX J0806.3+1527 (J0806 for short).
The nominal operation time of TianQin will be 5 yr, which is assumed throughout this paper.
There is expected to be only a very small drift of the detector orientation over a 5-yr period \cite{Ye:2019txh}, and, since that 
should have negligible effect on the present study,  we will not consider it.
The location of J0806 is close to the ecliptic plane, so the constellation plane of TianQin will be 
nearly perpendicular to the ecliptic plane.
The TianQin satellites will have nearly identical orbits and nearly identical periods, which will be about $T=3.6$ days \cite{Luo:2015ght,Hu:2017,Hu:2019}.

In the Solar Ecliptic Coordinate System where the x axis points towards the direction of the vernal equinox and the z axis is normal to and northward from the ecliptic plane, the location of the satellites at a given time can be formulated as \cite{Hu:2018yqb}
\bea \label{eq:orbit}
x(t)&=&R\cos\alpha_e+\frac{1}{2}R\cdot e_e\cdot \cos(2\alpha_e-3)\nn\\
&&+\frac{1}{\sqrt{3}}L\cdot (\cos\theta\cos\phi\cos\gamma_e-\sin{\phi}\sin{\gamma_e})\,,\nn\\	
y(t)&=&R\sin\alpha_e+\frac{1}{2}R\cdot e_e\cdot \sin(2\alpha_e)\nn\\
&&+\frac{1}{\sqrt{3}}L\cdot (\cos\theta\sin\phi\cos\gamma_e+\cos{\phi}\sin{\gamma_e})\,,\nn\\
z(t)&=&-\frac{1}{\sqrt{3}}L\cdot \sin\theta\cos\gamma_e\,,
\eea
where $R=1$AU is the semimajor axis, $e_e=0.0167$ is the eccentricity, and $\alpha_e=2\pi{f_e}t+\Phi_e$ is the phase, 
with $f_e=1/$yr and $\Phi_e$ being, respectively, the frequency and initial phase of the orbit of  Earth. The angular parameters ($\theta$, $\phi$) specify the space direction to J0806, and $\gamma_e=2\pi{t}/T+2\pi{n}/3+\delta$ $(n=1,2,3)$ are the phases of the satellites in their geocentric orbits, where $\delta$ is some reference phase that can be set to zero.

The effect of a propagating \ac{GW} ${h}(\xi)$ on the optical path length $L_{ij}$ starting from the satellite $i$ at time $t-L_{ij}$ and arriving at the satellite $j$ at time $t$ can be expressed as \cite{Cornish:2001qi,Rubbo:2003ap}
\be \label{eq:OpticalPath}
\delta{L}_{ij}(t)=\frac{1}{2}\frac{\hat{r}_{ij}(t)\otimes\hat{r}_{ij}(t)}{1-\hat{k} \cdot\hat{r}_{ij}(t)}:\int^{\xi_j}_{\xi_i}{h}(\xi)d\xi\,,
\ee
where $\xi_i$ is the phase and $\hat{r}_{ij}(t)$ is the unit vector from the satellite $i$ to the satellite $j$ at time $t$. Two independent Michelson interferometer signals can be constructed \cite{Cutler:1997ta}:
\bea
h_1(t)&=&[\delta{L_{12}}(t)-\delta{L_{13}}(t)]/L\,,\nn\\
h_{2}(t)&=&\frac{1}{\sqrt{3}}[\delta{L_{12}}(t)+\delta{L_{13}}(t)-\delta{L_{23}}(t)]/L\,.
\eea

Note that the orbital motion of the TianQin satellites also contributes a phase modulation (due to the Doppler shift) to the observed signal:
\be
\Phi^D(t)=2\pi\nu(t)R\sin\theta_S\cos[\phi(t)-\phi_S]\,,
\ee
where $2\nu$ is the frequency of the \ac{GW} signal, ($\theta_S$, $\phi_S$) are the angular coordinates of the source, and $\phi(t)=\alpha_e$ is the orbit phase.

\subsection{Detector noise}\label{sub:noise}

\begin{figure}
\centering
\includegraphics[width=\columnwidth,clip=true,angle=0]{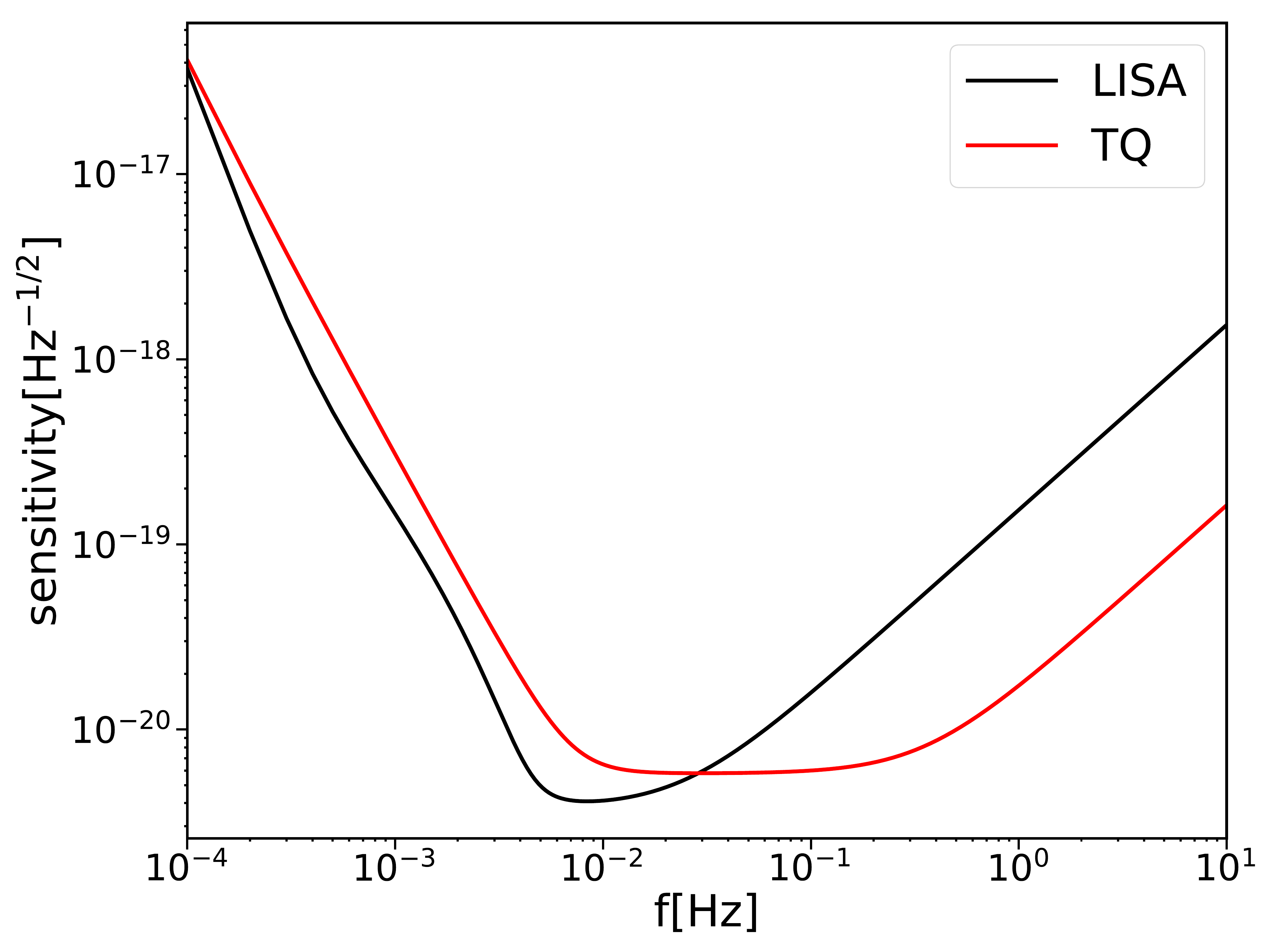}
\caption{The sensitivity curve for TianQin and LISA.}
\label{fig:asd}
\end{figure}

The noise model of TianQin is encoded in the following  sensitivity curve \cite{Huang:2020rjf,Wang:2019ryf,Shi:2019hqa,Finn:1992wt}:
\bea
\label{noise}
  S_n(f) &=&\frac{1}{L^2}\left[\frac{4S_a}{(2\pi f)^4}\left(1+\frac{10^{-4} {\mathrm Hz}}{f}\right) +S_x\right]\nn\\
  &&\times\left[ 1+0.6\left(\frac{f}{f_*}\right)^2 \right]\,,
\eea
where $S_a^{1/2}=1\times10^{-15}$m s$^{-2}/$Hz$^{1/2}$ characterizes the residual acceleration on a test mass playing the role of an inertial reference, $S_x^{1/2}=1\times10^{-12}$m/Hz$^{1/2}$ characterizes the one-way noise of the displacement measurement with intersatellite laser interferometry, and $f_*=1/(2\pi L)$ is the transfer frequency \cite{Luo:2015ght}.
An illustration of the sensitivity curve of TianQin is given in Fig. \ref{fig:asd}. For comparison, we also plot the LISA sensitivity curve in Fig. \ref{fig:asd} based on Ref. \cite{Babak:2017tow}.

At the low-frequency end of the TianQin observation range, there exist numerous compact binaries in the Galaxy, whose \ac{GW} signals overlap  to give rise to a stochastic unresolved signal, sometimes referred to as the foreground.
Preliminary analyses suggest that such a foreground will be consistently below the sensitivity curve  for resolved sources, given that the operation time of TianQin is limited to 5 yr \cite{Liang:2019}.  We therefore do not consider the effect of Galactic compact binaries throughout this work.

\section{Method}\label{sec:method}

\subsection{Signal-to-noise ratio}\label{sec:snr}


In order to study the prospects of detecting EMRIs with TianQin, one can calculate the \ac{SNR}. Previous studies have shown that \acp{EMRI} with \acp{SNR} as low as 15 can be detected by LISA under favorable circumstances \cite{Babak:2009cj}.
In this paper, we adopt the more conservative \ac{SNR} threshold of 20 for detection, following Ref. \cite{Babak:2017tow}.

Using the noise-weighted inner product between two signals $s_1(t)$ and $s_2(t)$,
\be
\label{eq:inner}
(s_1|s_2)=2\int_{0}^{\infty}\frac{\tilde{s}_1(f)\tilde{s}_2^\ast(f)+\tilde{s}_1^\ast(f) \tilde{s}_2(f)}{S_n(f)}df\,,
\ee
where $\tilde{s}_i(f)$, $i=1,2,$ are the Fourier transforms of $s_i(t)$, the \ac{SNR} can be defined as
\be
\label{eq:snr}
\rho=(h|h)^{1/2}=2\Big[\int_{0}^{\infty}\frac{\tilde{h}(f)\tilde{h}^\ast(f)}{S_n(f)}df\Big]^{1/2}\,,
\ee
where $h(t)$ is the \ac{GW}-induced signal in the detector. The Fourier transform $\tilde{h}(f)$ is obtained from $h(t)$ by applying a discrete Fourier transform:
\be
\tilde{h}\left(\frac{k}{N\Delta t}\right)=\Delta{t}\sum^{N}_{n=1}h(n\Delta t)e^{-i2\pi{k}n/N},
\ee
where $\Delta{t}$ is the sampling interval.

The basic TianQin data stream consists of data segments each lasting for 3 months for protection from heat instability, and the total accrued \ac{SNR} is obtained as the root sum square of the individual \ac{SNR} from each data segment. The same root-sum-square rule is also used when combining the contributions from different interferometer signals (when we consider TianQin operating within a detector network).

\subsection{Fisher information matrix}\label{fim}


%

The existence of noise leads to uncertainties in the inference on source parameters.
To quantify these uncertainties, we use the \ac{FIM} method to obtain the lowest-order expansion of the posteriors (valid in the high \ac{SNR} limit), which can be more accurately estimated through a full Bayesian parameter estimation analysis.
Indeed, we note that the \ac{FIM} method can be used as a fast assessment of the expected parameter estimation capabilities of an experiment, but the obtained $\Sigma$  represents only the Cramer-Rao bound of the covariance matrix. More advanced techniques are required to obtain more realistic results \cite{Vallisneri:2007ev,Rodriguez:2013mla}.

The \ac{FIM} is defined as
\be
\Gamma_{ij}=\Big(\frac{\partial\tilde{h}(f)}{\partial{\theta^i}}\Big| \frac{\partial\tilde{h}(f)}{\partial{\theta^j}}\Big)\,,
\ee
where $\theta^i$, $i=1,2,...,$ are the parameters appearing in the template $\tilde{h}(f)$.
When  multiple interferometers are present, the network's \ac{FIM} can be obtained as the sum of the individual \ac{FIM} from each interferometer.

The \ac{EMRI} waveform, even assuming the relatively simple AK model, is rather complicated, and it is difficult to obtain general analytical expressions for the partial derivatives $\partial{h(f)}/\partial{\theta^i}$. We therefore approximate derivatives with respect to the parameters by numerical  finite differences. In the lowest-order expansion (i.e. in the high \ac{SNR} limit), the variance-covariance matrix can be obtained as the inverse of the \ac{FIM}:
\be
\Sigma_{ij}\equiv\langle \delta\theta_i\delta\theta_j\rangle=(\Gamma^{-1})_{ij}.
\ee
From the variance-covariance matrix, the uncertainty $\sigma_i$ of the $i {\rm th}$ parameter $\theta_i$ can be obtained as
\be
\sigma_i=\Sigma_{ii}^{1/2}\,.
\ee
We also note that it is often meaningful to discuss the sky localization in terms of the solid angle $\Delta\Omega$
corresponding to the  error ellipse for which there is a probability $\exp(-1)$ for the source to be outside of it~\cite{Babak:2017tow}, which can be expressed as a combination of the uncertainties on the ecliptic longitude angle $\phi_S$ and the ecliptic latitude angle $\theta_S$,
\be
\Delta\Omega=2\pi\big|\sin \theta_S\big|\sqrt{\Sigma_{\theta_S}\Sigma_{\phi_S}-\Sigma_{\theta_S \phi_S}^2}\,.
\ee

\section{Results}\label{sec:result}

\subsection{Horizon distance}\label{HD}

\begin{figure*}[!htbp]
	\centering
\includegraphics[width=.6\textwidth]{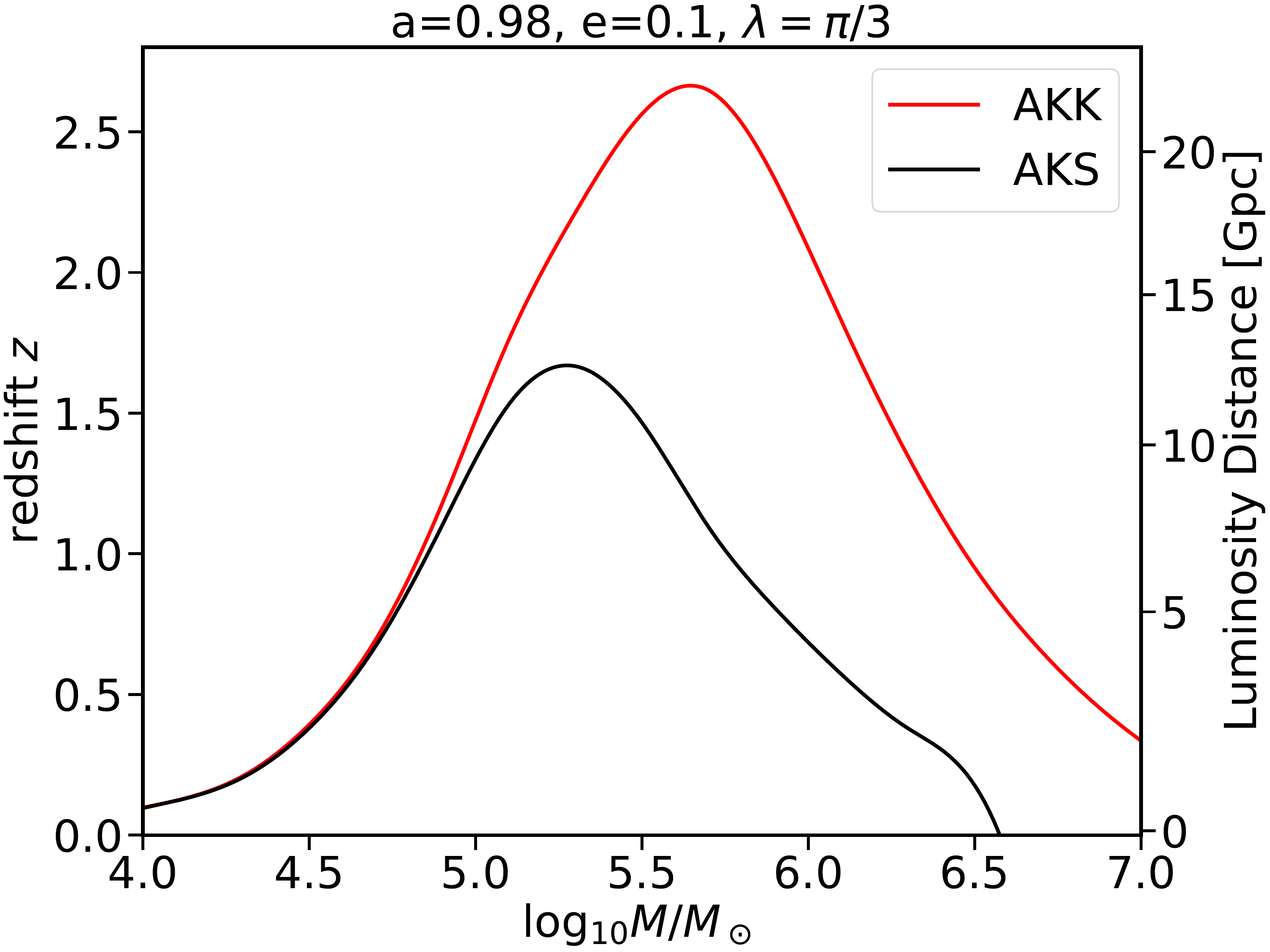}
\caption{TianQin's horizon distance for \ac{EMRI} systems as a function of the \ac{MBH} mass, assuming a \ac{MBH} spin of 0.98, eccentricity of 0.1 and inclination angle of $\pi/3$. The black line adopts the AKS waveform, while the red curve adopts the AKK waveform.}\label{fig:hordis}
\end{figure*}

As a first assessment of TianQin's capability of detecting \acp{EMRI}, we compute the horizon distance, i.e., the farthest distance at which
an \ac{EMRI} source can be detected, or, equivalently, the farthest distance at which
the SNR exceeds our detection threshold of 20, under the most favorable conditions possible \cite{Chen:2017wpg}.

For the seven intrinsic parameters $m$, $M$, $S/M^2$, $e_0$, $\tilde{\gamma_0}$, $\phi_0$, and $\lambda$, we fix the mass of \ac{CO}  to $m=10 M_\odot$ 
the orbital eccentricity to $e_{\rm lso}=0.1$, the \ac{MBH}'s spin to $a=0.98$, and the inclination angle to $\lambda=\pi/3$, while the initial condition for $\tilde{\gamma_0}$ is set to 0, although that choice has a marginal effect on the \ac{SNR}.
One can see from Eq.~(\ref{eq:OpticalPath}) that TianQin has the strongest response to sources sitting on the line passing through the detector and J0806, so the source is placed in the direction of J0806. We also fix $\Phi_0, \alpha_0$ to 0 and  $\theta_K, \phi_K$ to $\pi/4$,while the plunge time is taken to be 5 yr, which is the mission time of TianQin. These  values of the parameters are all set at the moment when the \ac{CO} reaches the Kerr LSO of the \ac{MBH}.

The maximum redshift at which \acp{EMRI} can be detected by TianQin with a threshold \ac{SNR} of 20 is illustrated in Fig. \ref{fig:hordis} as a function
of the MBH mass. The red curve (corresponding to AKK waveforms) would allow for a larger detection range and features better sensitivity to systems with heavier \acp{MBH} than the black curve (which more conservatively uses AKS waveforms). 
	This is mainly due to the fact that, by adopting a cutoff at later times (i.e., higher frequencies), the AKK waveforms include the larger GW amplitudes emitted when the \ac{CO} is very close to the \ac{MBH}.  Thus, farther events are expected to be detectable under the fixed \ac{SNR} threshold. 
	The maximum horizon distance for  AKK waveforms corresponds to $ z\approx2.6$, and to a \ac{MBH}  mass around $4\times 10^5 M_\odot$.
	For the AKS waveform, on the other hand, the maximum horizon distance is smaller, with a corresponding redshift of about 1.6.
	The  \ac{MBH} mass for which the AKS horizon distance peaks is also smaller and around $2\times 10^5 M_\odot$. 
	This feature can also be explained in a simple  manner: For the same \ac{EMRI} system, AKK waveforms extend to higher frequencies,
       but that high-frequency component is most important for high-mass systems, whose low-frequency inspiral
produces little SNR as it lies at lower frequencies than TianQin's sensitivity sweet spot. Therefore, when using AKS waveforms (for which
that high-frequency part is absent), the SNR of high-mass EMRIs is suppressed.

\subsection{Detection rate}\label{sec:DetRate}

\begin{table*}[!htbp]
\caption{The expected detection rate of \acp{EMRI} with TianQin for different astrophysical models.   The physical assumptions of the 12 models are described in Table I of Ref. \cite{Babak:2017tow}. The numbers are broken into different \ac{MBH} mass ranges in the middle three columns, $M_{10}\equiv\log_{10}(M/M_\odot)$, while the rightmost column summarizes the total detection rate. Numbers in brackets correspond to detection rates with AKS waveforms, while numbers outside brackets assume the AKK waveform.}\label{tab:detnum}
\begin{center}
\setlength{\tabcolsep}{7.5mm}
\begin{tabular}{|c|c|c|c|c|c|}
\hline
\multirow{2}{*}{Model}&\multirow{2}{*}{event rate (yr$^{-1}$)}&\multicolumn{3}{c|}{Detection rate of TianQin in mass range (yr$^{-1}$)}&\multirow{2}{*}{Total (yr$^{-1}$)}\\
\cline{3-5}
&&$M_{10}< 5$&$5 < M_{10}  < 6$&$6 < M_{10}$&\\\hline
M1&1600&1(1)&25(11)&8 (1)&34 (13)\\
M2&1400&0(0)&18(12)&2(0)&20 (12)\\
M3&2770&0(0)&83(28)&27 (2)&110(30)\\
M4&520&1 (0)&42(28)&7(3)&49(31)\\
M5&140&0(0)&4(2)&4(0)&8(2)\\
M6&2080&1(1)&40(22)&23(0)&64(23)\\
M7&15800&18(18)&187(121)&55(4)&260(143)\\
M8&180&0(0)&5(0)&1(0)&6(0)\\
M9&1530&2(1)&16(14)&2(1)&20(16)\\
M10&1520&0(0)&18(14)&0(0)&18(14)\\
M11&13&0(0)&0(0)&0(0)&0(0)\\
M12&20000&13(11)&273(113)&150(2)&436(126)\\\hline
\end{tabular}
\end{center}
\end{table*}

We compute the expected detection rates for \ac{EMRI} systems with TianQin by using the 12 astrophysical models developed in Ref. \cite{Babak:2017tow} and reviewed in Sec.
 \ref{subsec:astro}.
For each of the 12 models, we construct  catalogs of simulated events with both the number of events and their physical parameters randomly generated according to the underlying distribution.
Five parameters, including $M, m, a, \lambda, and z$, are inherited from the catalog realizations used by Ref. \cite{Babak:2017tow}, while all other parameters are randomly extracted again. In more detail,
the plunge time is distributed uniformly within the mission lifetime of TianQin (5 yr).
The sky positions of the sources ($\theta_S, \phi_s$) and their spin orientations  ($\theta_k,\phi_k$) are drawn from an isotropic distribution on the sphere.
The phase parameters $\Phi, \tilde{\gamma}, and \alpha$ at plunge  are uniformly distributed in $[0,2\pi]$.
The orbital eccentricity at plunge is drawn from a uniform distribution in $[0,0.2]$.

The \ac{SNR} for all events is calculated, and events with \ac{SNR} larger than 20 are considered as detected.
Again, we perform calculations with  AKS and  AKK waveforms separately (cf. Sec. \ref{sub:wave}). 
For both  AKK and AKS waveforms, the physical parameters are assumed to be measured at the Schwarzschild LSO.
In both cases, each system is then  evolved backward to a sufficiently long time before merger. 

We present the expected detection rates for different models in Table \ref{tab:detnum}, where the results for AKS (AKK) waveforms are inside (outside) the brackets.
The overall detection rates are summarized in the rightmost column, and a breakdown of the rates for three different \ac{MBH} mass ranges, $M_{10}<5$, $5<M_{10}<6$, and $M_{10}>6$ with $M_{10}\equiv\log(M/M_\odot)$, is also given.

For most of the 12 models, the expected detection rates vary from dozens to hundreds of events per year, regardless of whether AKS or AKK waveforms are used.
Models M5, M8, and M11 predict, however, significantly smaller detection rates, mainly because of the significantly lower  intrinsic \ac{EMRI} rate in the models themselves.
Note that   using the AKK waveforms generally predicts a significantly higher detection rate than using  AKS waveforms.
This happens because sources with prograde orbits are about 40\% more numerous than those  with retrograde orbits. Sources on
prograde orbits can have the Kerr LSO closer to the MBH, and, thus, AKK waveforms accrue  much higher \acp{SNR}.

One can also see from Table \ref{tab:detnum} and Fig. \ref{fig:hordis} that the majority of \acp{EMRI} exceeding the \ac{SNR} detection threshold are those with masses $10^5-10^6M_\odot$.
A similar feature was also found to hold  for the detectability  of massive binary black hole mergers using TianQin \cite{Wang:2019ryf}.
This feature is mostly related to the frequency dependence of the sensitivity curve and to the relation between the \ac{MBH} mass and the peak frequency of a \ac{GW} signal.

We remark that the calculations performed in this section
account for EMRIs and not for direct plunges into the MBH. In fact,
as mentioned in Sec. \ref{subsec:astro}, for each EMRI there is expected to
be a potentially sizable number $N_p$ of COs plunging directly into the MBH along low angular momentum orbits. We have, however, verified
that these plunges typically have a SNR as high as a few \cite{Berry:2013poa,Han:2020dql}, and are, thus, not easily detectable by TianQin. 

\subsection{Parameter estimation}

\begin{figure*}[htbp]
	\begin{subfigure}[b]{0.48\linewidth}
		\centering
		\includegraphics[height=6.0cm,width=7.1cm]{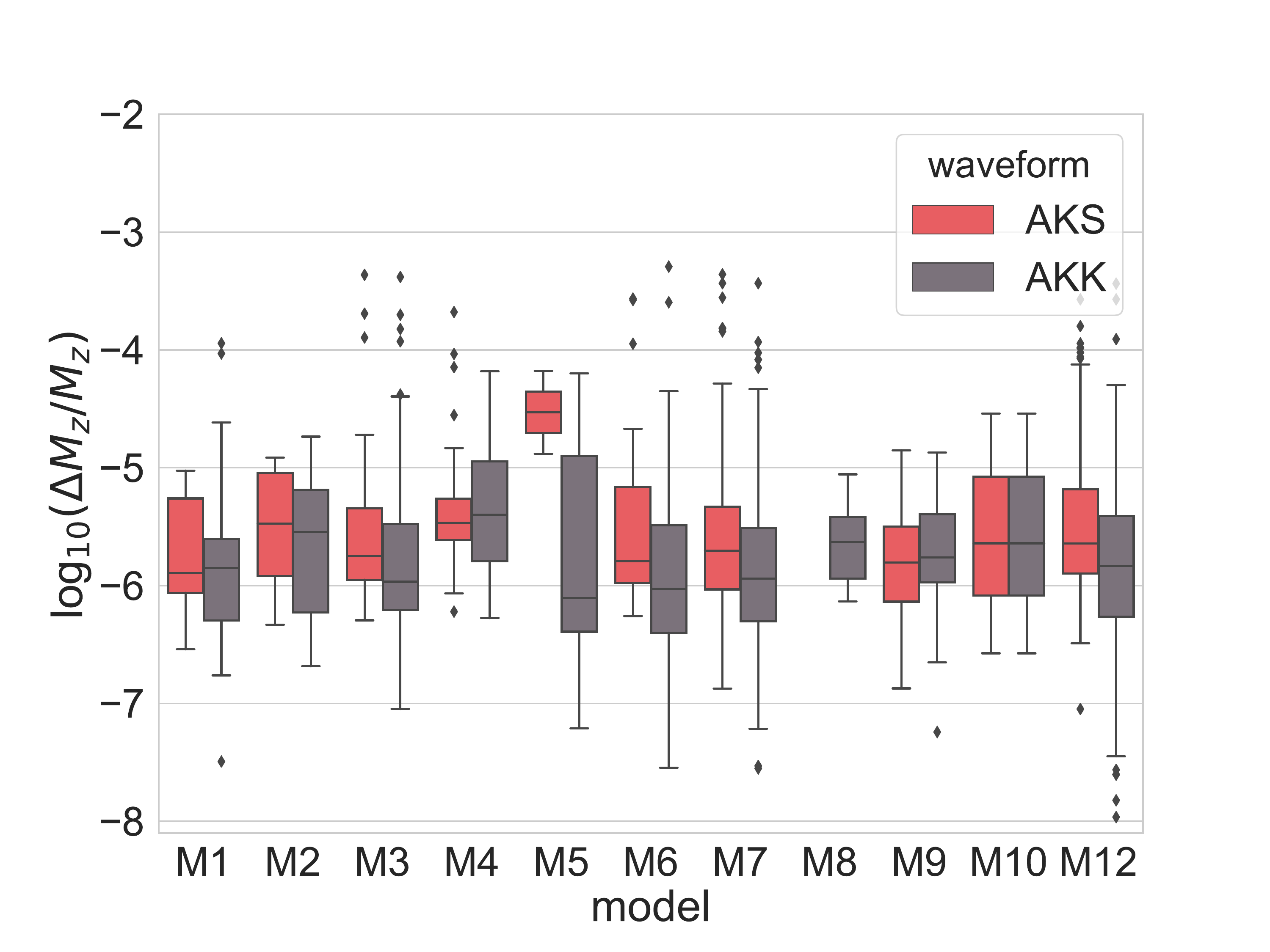}
	\end{subfigure}
	\begin{subfigure}[b]{0.48\linewidth}
		\centering
		\includegraphics[height=6.0cm,width=7.1cm]{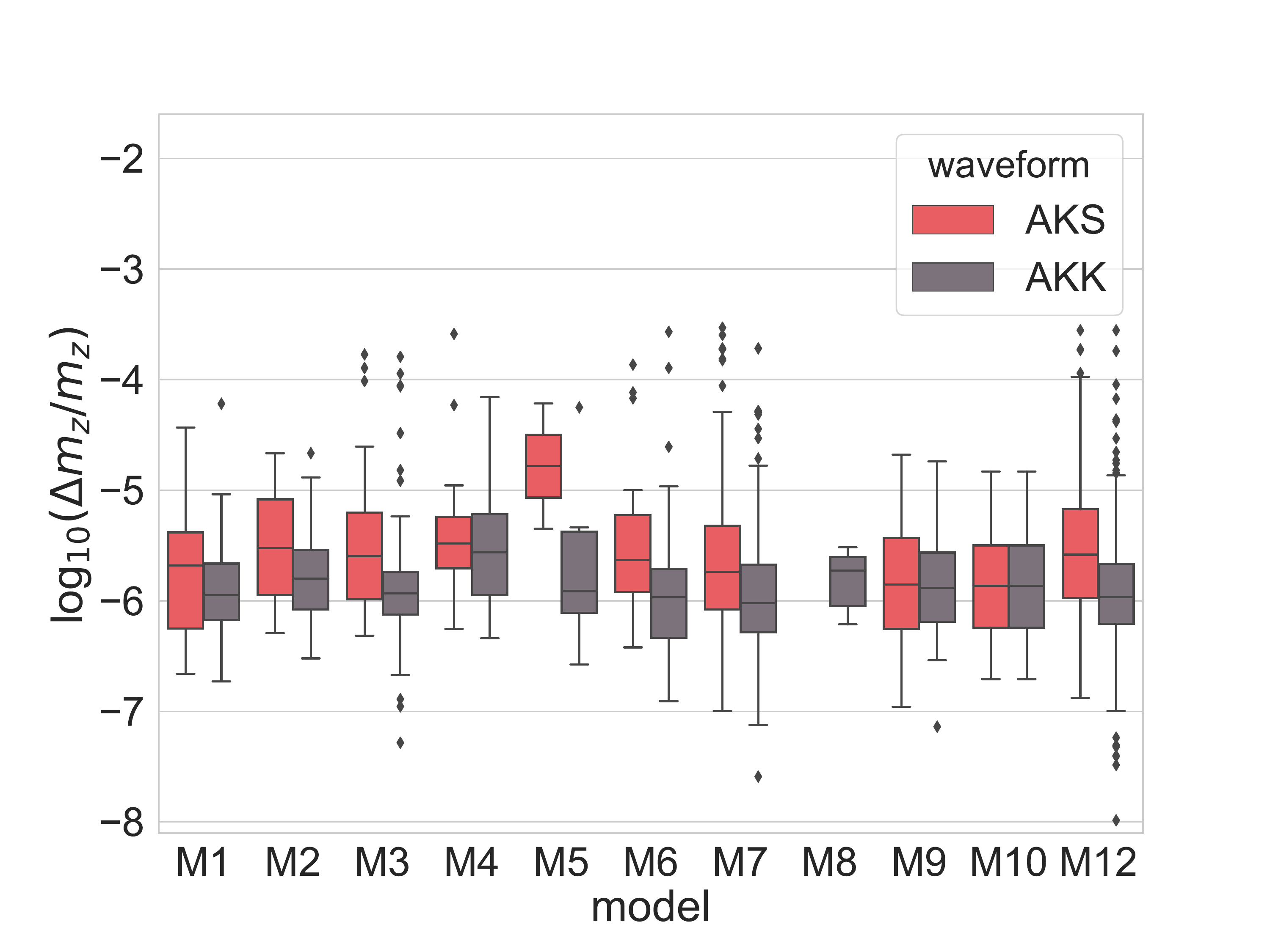}
	\end{subfigure}
	\begin{subfigure}[b]{0.48\linewidth}
		\centering
		\includegraphics[height=6.0cm,width=7.1cm]{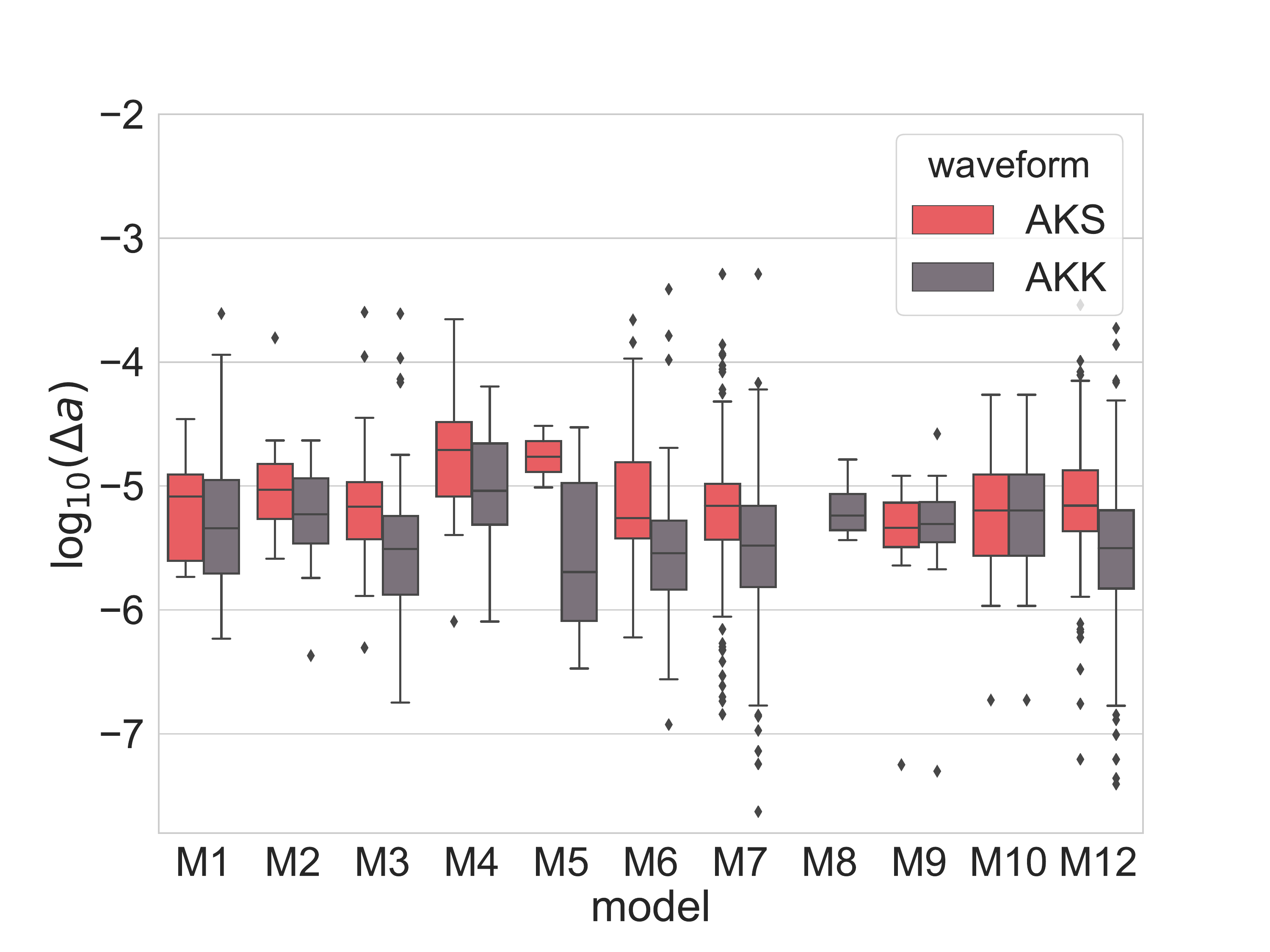}
	\end{subfigure}
	\begin{subfigure}[b]{0.48\linewidth}
		\centering
		\includegraphics[height=6.0cm,width=7.1cm]{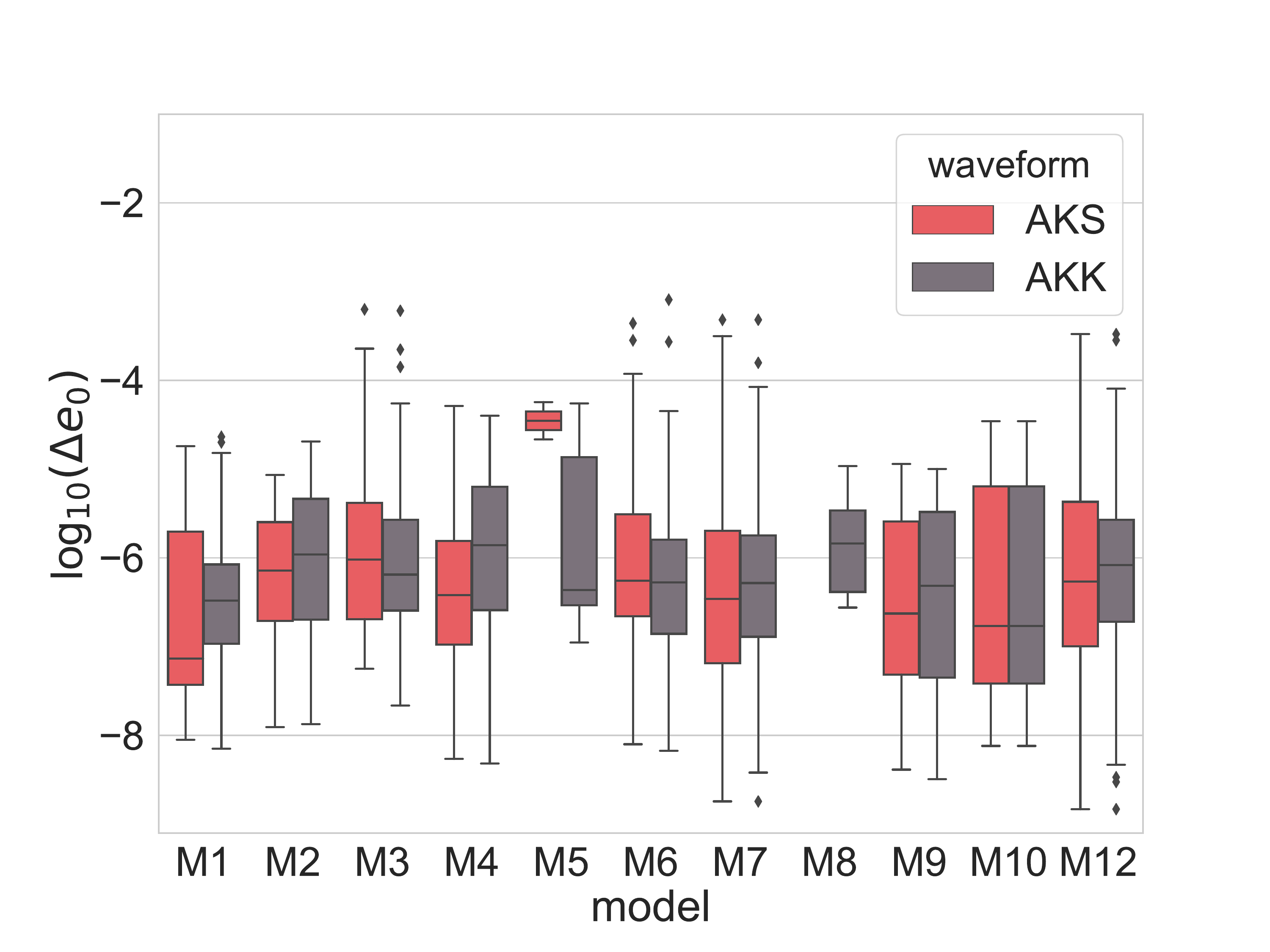}
	\end{subfigure}
	\begin{subfigure}[t]{0.48\linewidth}
		\centering
		\includegraphics[height=6.0cm,width=7.1cm]{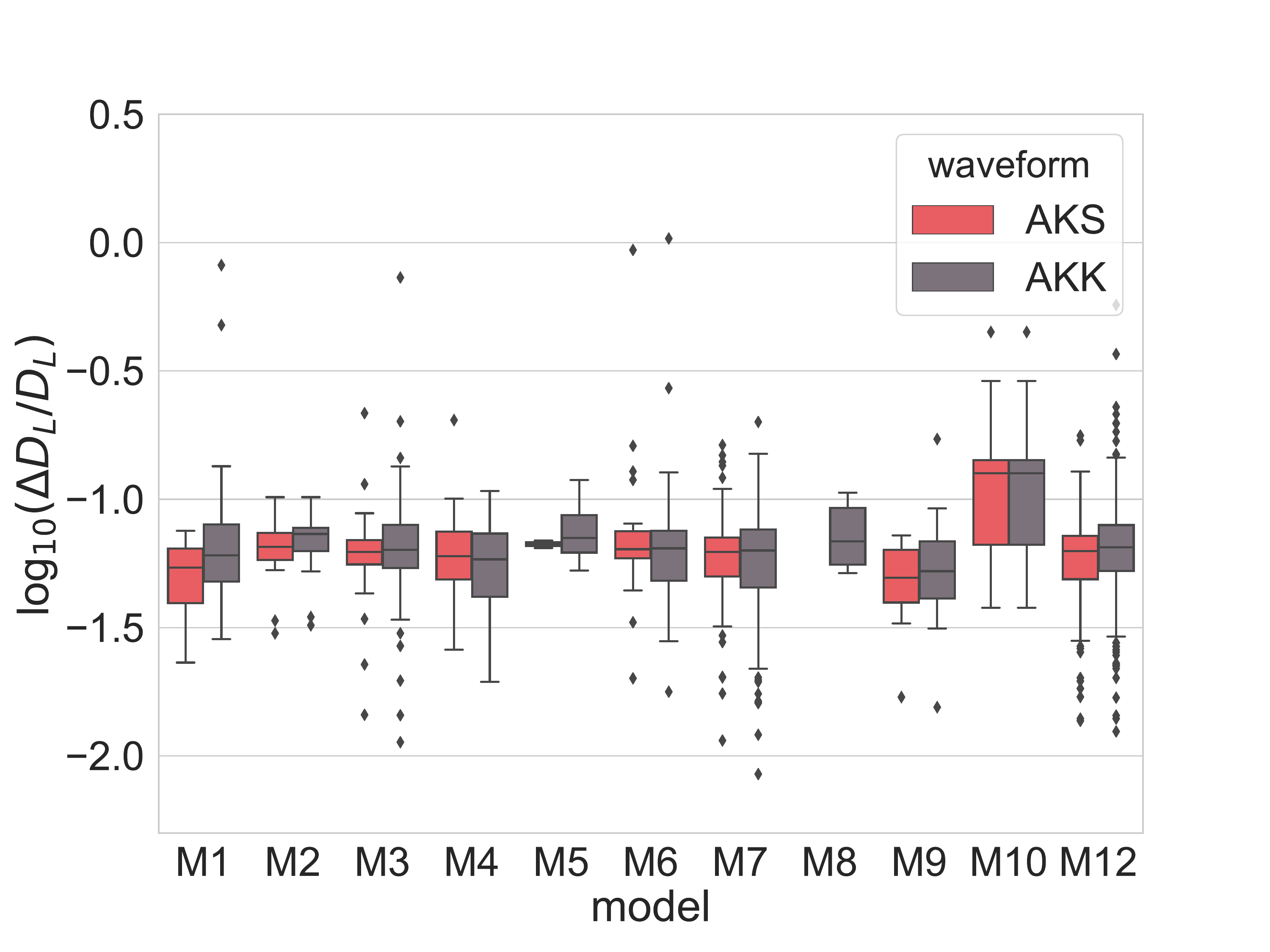}
	\end{subfigure}
	\begin{subfigure}[t]{0.48\linewidth}
		\centering
		\includegraphics[height=6.0cm,width=7.1cm]{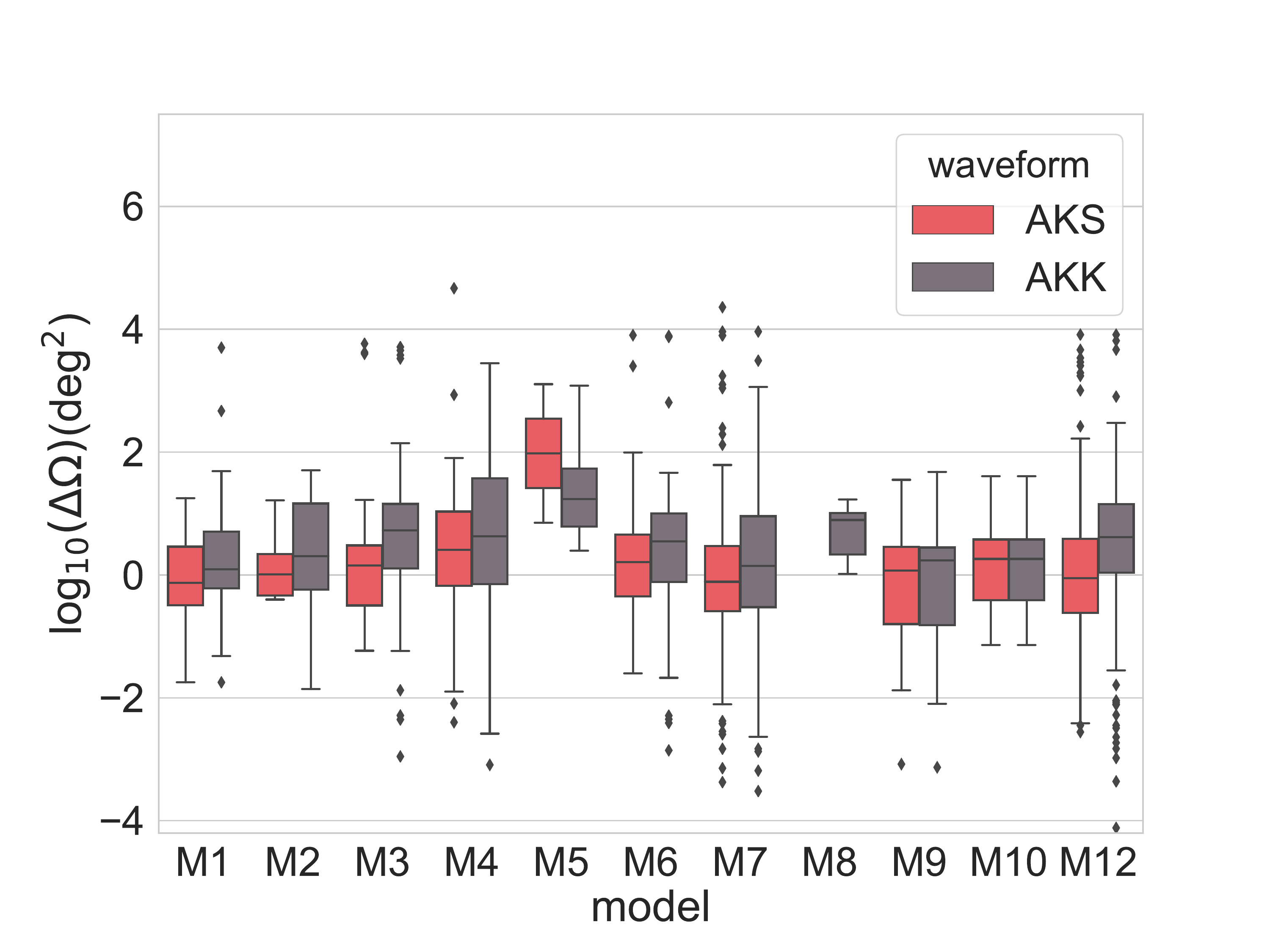}
	\end{subfigure}
        \caption{The parameter estimation precision for various astrophysical models, assuming AKS (red) and AKK (grey) waveforms respectively, shown in box plots. The interquartile range, or the central box,
        		represents the middle half of all values, with the median value also shown. The points outside the box are either covered by the whiskers when the most extreme point is no further than 1.5 times the box size or extreme points are plotted individually.}
	\label{Fig:aks}
\end{figure*}

As already mentioned, observation of the GW signal from EMRIs by space-based detectors may allow for 
testing the
 no-hair theorem \cite{Ryan:1995wh,Ryan:1997hg,Ryan:1997kh,Glampedakis:2005cf,Barack:2006pq,Gair:2012nm,Chua:2018yng}
and for revealing the possible presence of matter surrounding MBHs \cite{Barausse:2006vt,Barausse:2007dy,Gair:2010iv,Yunes:2011ws,Barausse:2014tra,Barausse:2014pra}. Moreover, EMRIs may
permit gaining information on
the mass distribution of MBHs \cite{Gair:2010yu} and on their host stellar
environments \cite{AmaroSeoane:2007aw}, as well as on  the expansion of the Universe \cite{MacLeod:2007jd}. All these goals, however, rely
on high-precision measurements of the source parameters. In this section, we therefore investigate the parameter estimation
of EMRIs with TianQin, using a FIM approach. 

Among the 14 parameters introduced in Sec. \ref{sub:wave}, one is generally mostly interested in the redshifted mass $m_z$ and the orbital eccentricity $e_0$ of the \ac{CO}, the redshifted mass $M_z$ and the spin $a$ of the \ac{MBH}, the luminosity distance to the source $D_L$, and the sky localization (the solid angle within which the source is located) $\Omega$.
The FIM-predicted uncertainties in the estimation of these parameters are given in Fig. \ref{Fig:aks}  for AKS and AKK waveforms  
 shown in the form of box plots. The central box represents the middle half with the central line indicating the median value. The whiskers covers the most extreme points when it is shorter than 1.5 times the box size; otherwise, the more extreme points are plotted individually.

\begin{figure*}[htbp]
   		\includegraphics[width=.6\textwidth]{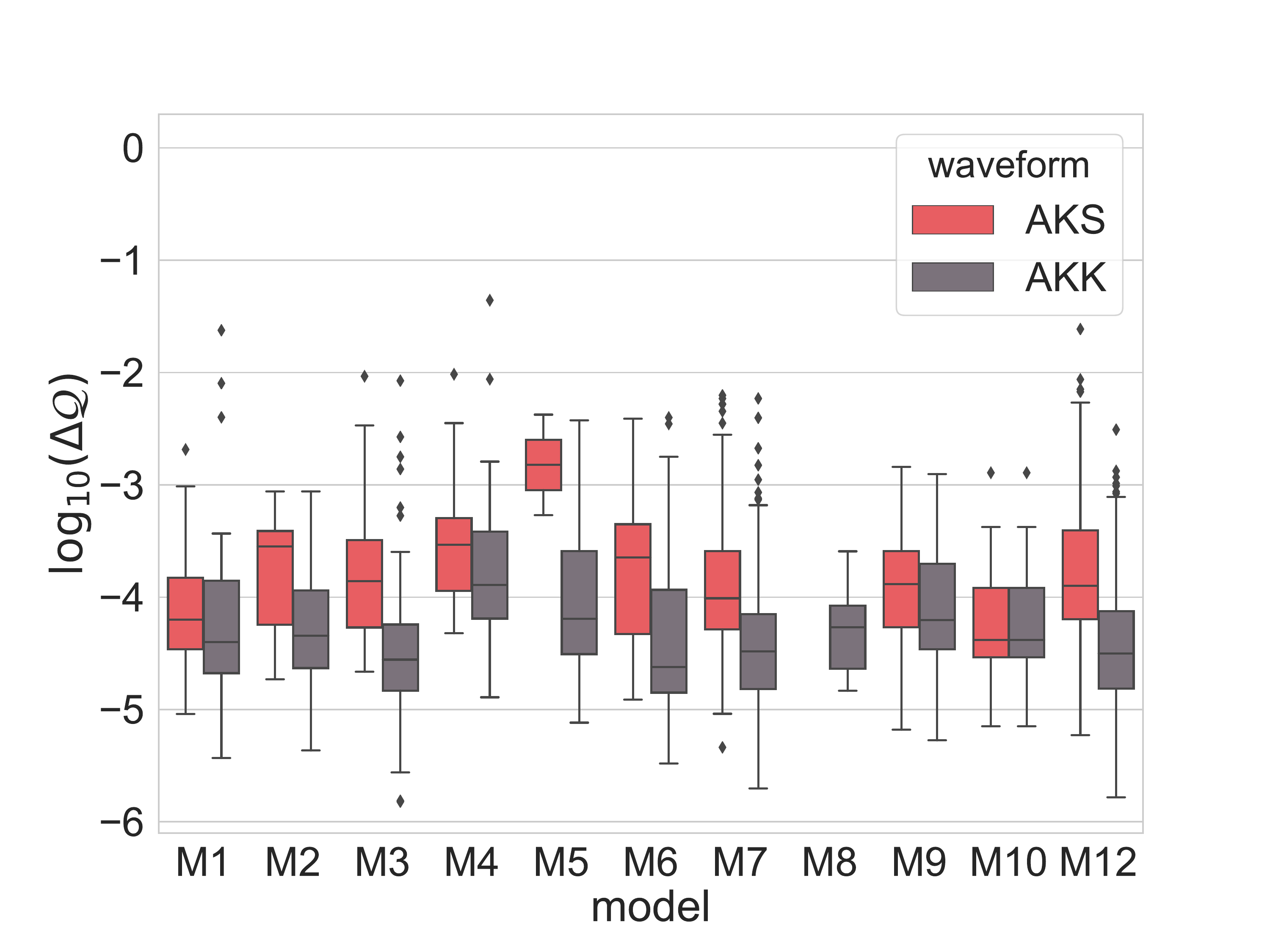}
        \caption{The parameter estimation precision for the ``anomalous'' quadrupole moment  $\mathcal{Q}$ defined in the text, for  AKS  (red) and AKK (gray) waveforms, shown in box plots, with the same setting as in Fig. \ref{Fig:aks}. }
   	\label{Fig:quad}
   \end{figure*}

We first notice that, although the 12 models cover a wide range of different astrophysical setups, the  distributions of the predicted errors are quite similar as expected from earlier studies conducted for the LISA mission \cite{Babak:2017tow}. Although notice that, for the models M8 and M11, the ratio between plunges to EMRIs is extremely high, so that the expected event rates are among the lowest, and, therefore, the detection rates can be as low as zero. Different choices of cutoff between the AKK and AKS waveforms can make a difference in the overall detectability in model M8, where we expect to detect six events assuming AKK waveform, no detection is expected when adopting AKS waveform. This leads to the difference in Figs. \ref{Fig:aks} and \ref{Fig:quad}.
%
There is also an intriguing similarity between the results of AKS and AKK, as one may have expected much better results for AKK waveforms.
Indeed, for a prograde source that can be detected with both AKS and AKK waveforms, the precision of parameter estimation is certainly better with AKK waveforms, because the \ac{SNR} is greater.
However, there is a large portion of events that can be detected with  AKK waveforms, but which do not have enough \ac{SNR} when using AKS waveforms.
 The high precision achieved for the high \ac{SNR} events is largely diluted by these relatively low \ac{SNR} events.
In this sense, the similarity between AKS and AKK is a demonstration of the strong link between the \ac{SNR} and the precision of parameter estimation.

Figs. \ref{Fig:aks}  also shows a stark difference between  intrinsic and  extrinsic parameters.
The intrinsic parameters are those that contribute to the phase of the \acp{GW}, such as the redshifted mass $M_z$ and the spin $a$ of the \ac{MBH}, and the redshifted mass $m_z$ and the orbit eccentricity $e_0$ of the \ac{CO}. Parameters such as the luminosity distance $D_L$ and the sky localization $\Omega$  affect only the amplitude of the \acp{GW}, and are referred to as extrinsic.
Assuming a typical observation time of $10^8$ s and a typical frequency of $10^{-2}$Hz, the number of wave cycles in an \ac{EMRI} signal is of the order of $10^6$.
Because of the huge number of cycles, a very slight change in the intrinsic parameters (and, hence in the phase) could change the cycle number by one, which is, in principle, detectable.
Therefore, a (relative) precision of the order of $10^{-6}$ is expected for the intrinsic parameters.
We indeed observe peaks roughly at this precision  in  Fig. \ref{Fig:aks}, for all models.
On the other hand, the estimation of the extrinsic parameters cannot benefit from the accumulation of a large number of wave cycles, and, therefore, the expected precision is much worse than for the intrinsic parameters.

We have also considered the possibility of testing the no-hair theorem by measuring the multipole moments
of the \ac{MBH} \cite{ryan1997accuracy,Barack:2006pq,Cornish:2001qi}.
A Kerr black hole satisfies the no-hair theorem and has a quadruple moment
$Q_K$ determined completely by its mass and spin: $Q_K=-a^2M^3$ \cite{hansen1974multipole}.
Here, we relax the Kerr hypothesis and allow for the quadrupole moment $Q$ to deviate from
the Kerr value. 
In Fig. \ref{Fig:quad}, we present the predicted errors on the dimensionless quantity $\mathcal{Q} \equiv (Q-Q_k)/M^3$, with
the two distributions corresponding to the AKS and AKK waveform for the 12 models, respectively.

\subsection{TianQin in a network of detectors}

The operation of TianQin inevitably leads to gaps between data or effectively lowering the duty cycle for the observation.
The proposed twin constellation configuration of TQ I+II, where two identical constellations differ only by orientation, has been considered in the following consideration.
The TQ II constellation would be perpendicular to both the ecliptic plane and the orbital plane of TQ.
Under such a design, and adopting the nominal 3 months operation followed by 3 months break, the TQ II can be configured in a relay mode so that when the TQ starts operation exactly when the TQ II ends,  there would be no gap between data and, therefore, significantly increase the duty cycle.
We briefly discuss the intriguing possibility that TianQin could be observing within a network of detectors, such as TQ I+II, TQ + LISA, and TQ I+II + LISA (see \cite{Huang:2020rjf} for a detailed explanation of each of the detector networks).

\begin{figure}
	\centering
	\includegraphics[width=\columnwidth,clip=true,angle=0]{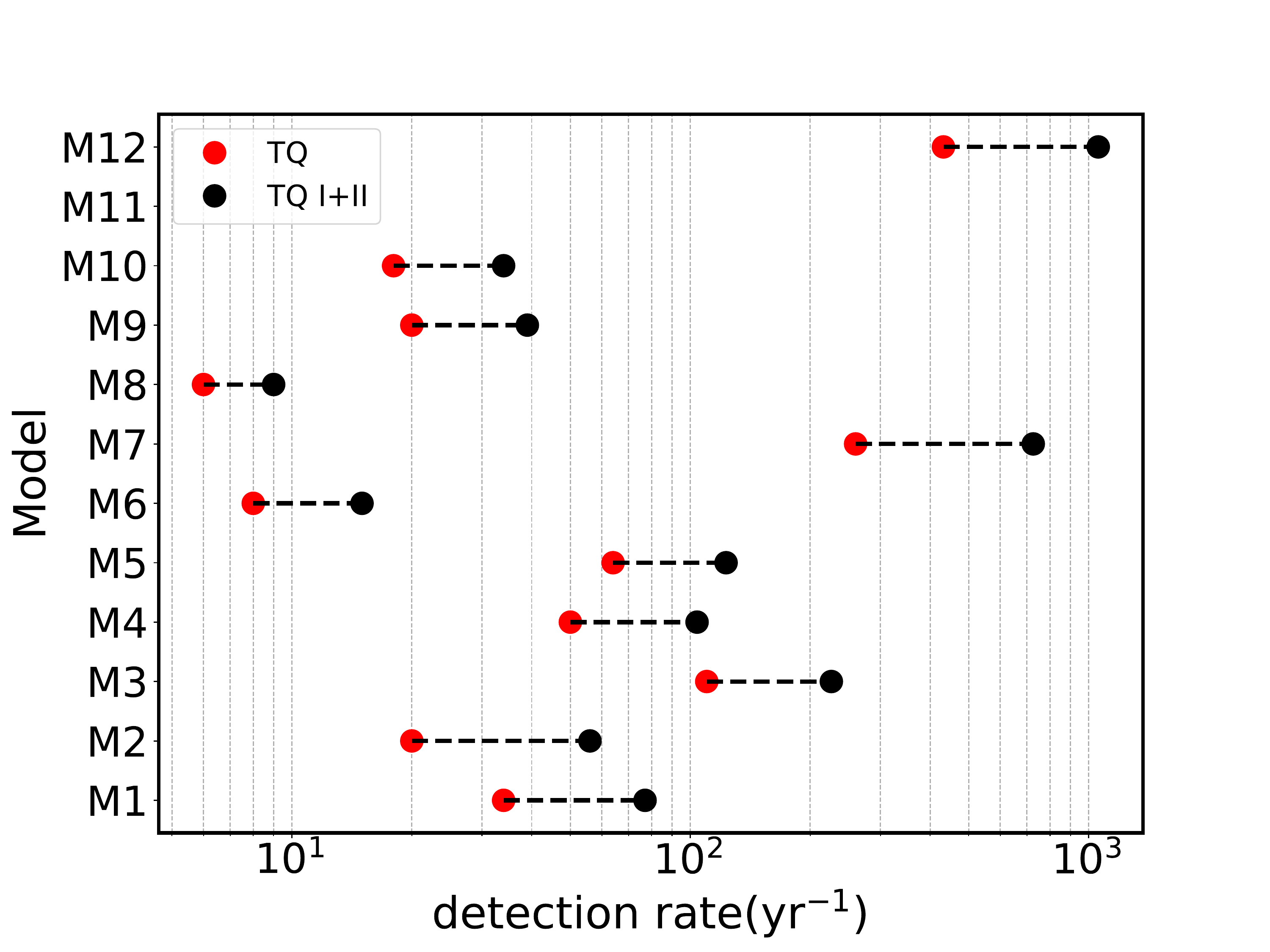}
	\caption{The expected detection rates of different models, using both the TQ  (red dots) and the TQ I+II (black dots) configuration.
		With a logarithm scale in detection rate, the length of the dashed line segments represents the factor of improvement of detection rates.}
	\label{fig:ratio}
\end{figure}

In Fig. \ref{fig:ratio}, we plot the expected detection rate using TQ  and TQ I+II, adopting AKK waveforms.
Note that, for burst signals, extending the observation time to full coverage would effectively double the detection rate.
However, since EMRIs are long-lived sources, the increase in the detection rate will not scale
with the duty cycle, and is shown in Fig. \ref{fig:ratio}. As can be seen, detection rates could increase
by a factor ranging from $\sim2$ in model 8 up to even $\sim 3$.

A comparison of the precision of parameter estimation with TQ, TQ I+II, LISA, TQ +LISA, and TQ I+II+LISA is given in Table \ref{tab:TQconstellation}.
In the calculation,  we considered EMRIs with four different MBH masses log$(M/M_\odot)=5, 5.5, 6, 6.5$ using the same set of parameters assumed for the horizon distance calculation, with the exception of  the plunge time, which is taken to be 1 yr, and the luminosity distance, which is taken to be 1 Gpc.
Notice that, due to the better sensitivity in lower frequency, LISA outperforms TianQin consistently, by no more than one order of magnitude.
We note that a detector network can usually improve on the precision of sky localization by more than 20 times, while for other parameters, the improvement can reach over 5 times.
As a comparison among the networks of detectors, TQ+LISA is consistently better than TQ I+II, while TQ I+II + LISA is always the best.

 \begin{table*}[!htbp]
 	\begin{center}
 		\caption{Parameter estimation precisions for different sources of TQ, TQ I+II, LISA, TQ+LISA and TQ I+II+LISA.}\label{tab:TQconstellation}
 		\setlength{\tabcolsep}{5.0mm}
 		\resizebox{\textwidth}{!}{\begin{tabular}{@{}cc|cccccc@{} }
 				\toprule
 				MBH mass& configuration&$\Delta{M_z}/M_z$& $\Delta{m_z}/m_z$& $\Delta{a}$& $\Delta{e_0}$& $\Delta{D_L}/D_L$& $\Delta\Omega$(deg$^2$)\\		
 				\hline
 				\multirow{4}{*}{log$(\frac{M}{M_\odot}) = 5.0$}&TQ&$2.25\times10^{-7}$&$9.15\times10^{-7}$&$1.46\times10^{-7}$&$1.71\times10^{-8}$&$2.23\times10^{-2}$&0.31\\
 				&TQ I+II&$1.54\times10^{-7}$&$7.3\times10^{-7}$&$1.15\times10^{-7}$&$1.25\times10^{-8}$&$2.19\times10^{-2}$&0.19\\
 				&LISA&$1.15\times10^{-7}$&$6.76\times10^{-7}$&$1.01\times10^{-7}$&$0.98\times10^{-8}$&$2.67\times10^{-2}$&0.18\\
 				&TQ+LISA&$1.01\times10^{-7}$&$4.86\times10^{-7}$&$0.80\times10^{-7}$&$0.84\times10^{-8}$&$1.70\times10^{-2}$&0.09\\
 				&TQ I+II+LISA&$0.91\times10^{-7}$&$4.67\times10^{-7}$&$0.75\times10^{-7}$&$0.76\times10^{-8}$&$1.69\times10^{-2}$&0.08\\			
 				\hline
 				\multirow{4}{*}{log$(\frac{M}{M_\odot}) = 5.5$}&TQ&$1.87\times10^{-6}$&$1.41\times10^{-6}$&$6.3\times10^{-7}$&$4.09\times10^{-7}$&$1.27\times10^{-2}$&$1.51$\\
 				&TQ I+II&$1.23\times10^{-6}$&$0.82\times10^{-6}$&$5.47\times10^{-7}$&$2.32\times10^{-7}$&$1.24\times10^{-2}$&0.46\\
 				&LISA&$0.97\times10^{-6}$&$0.68\times10^{-6}$&$5.06\times10^{-7}$&$2.16\times10^{-7}$&$1.15\times10^{-2}$&0.22\\
 				&TQ+LISA&$0.72\times10^{-6}$&$0.51\times10^{-6}$&$3.59\times10^{-7}$&$1.51\times10^{-7}$&$0.85\times10^{-2}$&$0.15$\\
 				&TQ I+II+LISA&$0.69\times10^{-6}$&$0.49\times10^{-6}$&$3.51\times10^{-7}$&$1.48\times10^{-7}$&$0.84\times10^{-2}$&$0.14$\\				
 				\hline
 				\multirow{4}{*}{log$(\frac{M}{M_\odot}) = 6.0$}&TQ&$6.63\times10^{-6}$&$3.53\times10^{-6}$&$9.66\times10^{-7}$&$5.53\times10^{-6}$&$9.7\times10^{-3}$&$4.88$\\
 				&TQ I+II&$3.08\times10^{-6}$&$1.87\times10^{-6}$&$6.30\times10^{-7}$&$2.55\times10^{-6}$&$9.11\times10^{-3}$&$1.61$\\
 				&LISA&$0.59\times10^{-6}$&$0.44\times10^{-6}$&$3.32\times10^{-7}$&$0.50\times10^{-6}$&$6.94\times10^{-3}$&0.19\\
 				&TQ+LISA&$0.57\times10^{-6}$&$0.40\times10^{-6}$&$2.74\times10^{-7}$&$0.48\times10^{-6}$&$5.48\times10^{-3}$&$0.16$\\
 				&TQ I+II+LISA&$0.57\times10^{-6}$&$0.40\times10^{-6}$&$2.72\times10^{-7}$&$0.48\times10^{-6}$&$5.45\times10^{-3}$&$0.15$\\ 				
 				\hline
 				\multirow{4}{*}{log$(\frac{M}{M_\odot}) = 6.5$}&TQ&$3.40\times10^{-6}$&$5.01\times10^{-6}$&$9.07\times10^{-7}$&$3.38\times10^{-6}$&$1.5\times10^{-2}$&$11.8$\\
 				&TQ I+II&$3.04\times10^{-6}$&$2.68\times10^{-6}$&$8.19\times10^{-7}$&$2.94\times10^{-6}$&$1.46\times10^{-2}$&$4.51$\\
 				&LISA&$0.54\times10^{-6}$&$0.89\times10^{-6}$&$2.74\times10^{-7}$&$0.68\times10^{-6}$&$0.62\times10^{-2}$&0.49\\
 				&TQ+LISA&$0.51\times10^{-6}$&$0.83\times10^{-6}$&$2.59\times10^{-7}$&$0.64\times10^{-6}$&$0.57\times10^{-2}$&$0.44$\\
 				&TQ I+II+LISA&$0.51\times10^{-6}$&$0.82\times10^{-6}$&$2.57\times10^{-7}$&$0.64\times10^{-6}$&$0.57\times10^{-2}$&$0.42$\\
 				\toprule
 			\end{tabular}}
 		\end{center}
 		\label{Table:paraEsII}
 	\end{table*}

\section{Summary and future work}\label{summary}

In this work, we have performed a preliminary study of the horizon distance, detection rate and precision of parameter estimation for \acp{EMRI} with TianQin. We have employed 12 astrophysical models encapsulating a wide range of different scenarios for the underlying \ac{EMRI} populations, 
which result in significantly different intrinsic EMRI rates (ranging from $\sim 10$ to $\sim 20000$ per year). Waveforms are described
by simple analytic kludge templates.

Adopting a detection threshold of \ac{SNR}=20, we find that most of the 12 astrophysical models predict that TianQin will detect dozens to thousands of \acp{EMRI}. The only model in which this is not the case (model 11) predicts that the vast majority of events should involve a \ac{CO} plunging directly into the \acp{MBH},
which results in very low rates irrespective of the GW detector configuration. 

As for the horizon distance, we find that EMRIs can be detected up to maximum redshifts varying from $1.6$ to $\sim 2.6$ according to
what waveform model is adopted (AKS versus AKK).  The MBH mass  yielding the maximum horizon distance also changes from around $2\times 10^5~\msun$ if AKS waveforms are used
to around $4\times 10^5~\msun$ for AKK waveforms. As a result, AKK waveforms  also predict larger
detection rates. Overall, this dependence on the waveform model highlights the need to
develop fast and accurate EMRI waveforms beyond the kludge approximation.

The expected precision of the parameter estimation is calculated using the \ac{FIM} method.
We find that the majority of detected events can determine the intrinsic parameters to within fractional errors of $\sim 10^{-6}$, while the errors on the extrinsic parameters are much less stringent.
	However, the majority of detected events can still determine the relative uncertainty in the luminosity distance with $10\%$ and the uncertainty in the sky localization to the level of about $10$ deg$^2$.
	The precise determination of the three-dimensional location might make it possible for \acp{EMRI} to be used as standard sirens for cosmology \cite{Schutz:1986gp,Tamanini:2016zlh,MacLeod:2007jd}, although further detailed studies are needed in this direction.

We briefly consider using EMRIs to put constraints on possible deviations from the Kerr quadruple moment, and we find the uncertainty in the dimensionless parameter ${\cal Q}$ peaks at about $\Delta {\cal Q} \sim10^{-4}$.

We also briefly consider the possible cases when TianQin is observing within a network of detectors. We find that such networks of detectors can improve the precision on sky localization by more than 20 times and the precision on other parameters as large as 5 times.

\begin{acknowledgments}
 This work has been supported by the Natural Science Foundation of China (Grants No. 11703098, No. 11805286, No. 11805287, No. 91636111, No. and 11690022), Guangdong Major Project of Basic and Applied Basic Research (Contract No. 2019B030302001), National Supercomputer Center in Guangzhou, European Union's H2020 ERC Consolidator Grant
``GRavity from Astrophysical to Microscopic Scales'' (Grant Agreement No. GRAMS-815673, to E.B.) and European Union's H2020 ERC Consolidator Grant
``Binary massive black hole astrophysics'' (Grant Agreement No. 818691 -- B Massive, to A.S.) .
 The authors also thank  Changfu Shi, Haitian Wang, Fu-Peng Zhang, Alvin J. K. Chua, Xian Chen, Stas Babak, and Jon Gair for helpful discussions.This work is supported by National Supercomputer Center in Guangzhou.
\end{acknowledgments}

\bibliographystyle{apsrev4-1}
\bibliography{reference}
\end{document}